\documentclass[10pt]{iopart}
\usepackage[colorlinks,bookmarks=false,citecolor=blue,linkcolor=red,urlcolor=blue]{hyperref}
\usepackage{graphicx,epstopdf,color}
\usepackage{amsfonts}
\usepackage{amsmath,amssymb,mathrsfs}
\usepackage{iopams}
\usepackage{bm}
\usepackage{pst-grad}

\def\bc{\begin{center}}
\def\ec{\end{center}}

\newcommand{\ket}[1]{\left|#1\right\rangle}

\newcommand{\up}{\uparrow}
\newcommand{\dw}{\downarrow}

\newcommand{\bea}{\begin{eqnarray}}
\newcommand{\eea}{\end{eqnarray}}

\def\ie{\emph{i.e.},\ }

\newcommand{\be}{\begin{equation}}
\newcommand{\ee}{\end{equation}}

\newcommand{\bfss}{{\boldsymbol{S}}}

\newcommand{\bfk}{{\boldsymbol{k}}}

\newcommand{\bfx}{{\boldsymbol{x}}}
\newcommand{\bfy}{{\boldsymbol{y}}}

\def\bmx{\begin{pmatrix}}
\def\emx{\end{pmatrix}}

\begin{document}

\title{Correlated Topological Phases and Exotic Magnetism with Ultracold Fermions}
%
\author{Peter P. Orth$^1$, Daniel Cocks$^2$, Stephan Rachel$^3$, Michael Buchhold$^4$, Karyn Le Hur$^5$ and Walter Hofstetter$^2$}
\address{$^1$ Institute for Theory of Condensed Matter, Karlsruhe Institute of Technology (KIT), 76131 Karlsruhe, Germany }
\address{$2$ Institut f\"ur Theoretische Physik, Goethe-Universit\"at, 60438 Frankfurt/Main, Germany}
\address{$^3$ Institute for Theoretical Physics, Dresden University of Technology, 01062 Dresden, Germany}
\address{$^4$ Institute for Theoretical Physics, University of Innsbruck, A-6020 Innsbruck, Austria}
\address{$^5$ Center for Theoretical Physics, Ecole Polytechnique, CNRS, 91128 Palaiseau Cedex, France }
\ead{peter.orth@kit.edu}

\begin{abstract}
Motivated by the recent progress in engineering artificial non-Abelian gauge fields for ultracold fermions in optical lattices, we investigate the time-reversal-invariant Hofstadter-Hubbard model. We include an additional staggered lattice potential and an artificial Rashba--type spin-orbit coupling term available in experiment. Without interactions, the system can be either a (semi)-metal, a normal or a topological insulator, and we present the non-Abelian generalization of the Hofstadter butterfly. Using a combination of real-space dynamical mean-field theory (RDMFT), analytical arguments, and Monte-Carlo simulations we study the effect of strong on-site interactions. We determine the interacting phase diagram, and discuss a scenario of an interaction-induced transition from normal to topological insulator. At half-filling and large interactions, the system is described by a quantum spin Hamiltonian, which exhibits exotic magnetic order due to the interplay of Rashba--type spin-orbit coupling and the artificial time-reversal-invariant magnetic field term. We determine the magnetic phase diagram: both for the itinerant model using RDMFT and for the corresponding spin model in the classical limit using Monte-Carlo simulations. 

\end{abstract}

\maketitle
\submitto{\jpb}

\section{Introduction and time-reversal-invariant Hofstadter-Hubbard model}
\label{sec:introduction}
Experiments with ultracold atoms in optical lattices can provide novel insight into traditional strongly correlated condensed matter phases in real materials~\cite{bloch:885,lewenstein_advphys_review_2007,Georges-FermiSummerSchoolLectures-2006}. Examples are the Mott-insulator to superfluid quantum phase transition in both the bosonic~\cite{greiner_nature_2002,spielman:080404,bakr_probing_2010} and the fermionic Hubbard model~\cite{PhysRevLett.94.080403, esslinger-FermionicMI-Nature-2008}, the physics of the BEC-BCS crossover between a Bose-Einstein condensate (BEC) of bosonic molecules and the Bardeen-Cooper-Schrieffer (BCS) pairing state of its fermionic constituents~\cite{PhysRevLett.92.040403, Greiner-BECToBCS-Nature-2003, PhysRevLett.93.050401, Jochim-BECToBCS-Science-2003, NozieresSchmittRink-BECToBCSCrossover-JLowTempPhys-1985}, or the emergence of quantum magnetic order in itinerant systems for large on-site interactions~\cite{Trotzky-SuperexchangeColdAtoms-Science-2008,PhysRevLett.91.090402,SimonGreiner-AFQuantumSpinChainColdAtom-Nature-2011}. Cold-atom lattice setups offer various advantages: they are free from disorder, can be experimentally probed with single-site precision (in two dimensions at least)~\cite{bakr_probing_2010, bakr_quantum_2009}, and they allow unique control over its system parameters such as lattice geometry or the nature of the constituent particles. Particularly attractive is the precise tunability of the interaction strength between the particles via Feshbach resonances, which allows for a quantitative study of the role of strong correlations in condensed matter systems. One central goal in the field is thus the investigation of possible high-temperature $d$-wave superfluidity in the 2D Fermi-Hubbard model~\cite{PhysRevLett.89.220407,0295-5075-87-6-60001, PhysRevLett.104.066406,PhysRevLett.106.035301,PhysRevA.86.023633, Hur20091452}. 

For a long time, however, the toolbox for quantum simulations was missing one important aspect: the equivalent of orbital magnetism. Since cold-atoms are charge neutral, they do not experience the Lorentz force due to external magnetic or electric fields. One way to simulate magnetism for cold atoms is to set the complete system in uniform rotation around a fixed axis~\cite{doi:10.1080/00018730802564122,RevModPhys.81.647}. This approach, however, is severely limited in experiment as it requires, for example, axial symmetry which is not present in optical lattice experiments. Recently, it was experimentally demonstrated that artificial gauge potentials for cold atoms may also be generated using laser beams~\cite{RevModPhys.83.1523,lin_synthetic_2009,lin_synthetic_2011}. The atom experiences an effective gauge field due to light-induced couplings between different atomic internal states. This leads to a dressed basis of states, and if the atom-light coupling is spatially modulated the moving atom picks up a geometric phase~\cite{berry-berryphase-1984,PhysRevLett.76.1788}. Importantly, this scheme is applicable in the presence of optical lattice potentials, where the hopping element acquires a phase factor from the well-known Peierls substitution~\cite{ruostekoski_particle_2002,jaksch_creation_2003,PhysRevLett.107.255301,PhysRevLett.108.225303, PhysRevLett.108.225304,cooper_optical_2011}. Furthermore, these artificial gauge fields may easily be generalized to the non-Abelian case or to represent spin-orbit interaction by coupling internal atomic levels that correspond to different pseudo-spin states of the atom~\cite{dudarev_spin-orbit_2004,osterloh_cold_2005, PhysRevLett.95.010404, PhysRevLett.97.216401, 0295-5075-80-2-20001, PhysRevA.77.043410, lin_spin-orbit-coupled_2011, PhysRevLett.109.095302, PhysRevLett.109.095301, PhysRevLett.109.145301}. The effective electromagnetic fields induced by these artificial gauge fields can be large, and in certain cases they can even give rise to the quantum (spin) Hall effect~\cite{girvin-prange_qheffect_book,PhysRevLett.95.146802,bernevig_quantum_2006,koenig_quantum_2007}, thus allowing for its study in a completely new experimental setting~\cite{PhysRevA.79.053639,  PhysRevA.79.063613, PhysRevA.81.033622, PhysRevA.82.013608, PhysRevLett.101.186807, PhysRevLett.105.255302,PhysRevA.83.023609, PhysRevA.85.033620,PhysRevLett.107.145301, PhysRevLett.100.070402, PhysRevA.84.063629, PhysRevLett.107.235301, PhysRevA.86.053804}. 


Here, we investigate the role of interactions in a spinful and time-reversal-invariant generalization of the paradigmatic Hofstadter model~\cite{PhysRevLett.105.255302, hofstadter_energy_1976} using real-space dynamical mean-field theory (RDMFT) and classical Monte-Carlo simulations~\cite{RevModPhys.68.13,1367-2630-10-9-093008,RevModPhys.80.395,RevModPhys.83.349}. We complement our numerical results with analytical arguments. This time-reversal-invariant \emph{Hofstadter-Hubbard model}~\cite{PhysRevLett.109.205303} can be realized using cold-atoms in artificial gauge fields~\cite{PhysRevLett.105.255302,gerbier_gauge_2010}. It hosts a number of fundamental physical problems that we address below such as the stability of topological phases with respect to interactions and the development of exotic magnetic order at large interactions due to spin-orbit coupling.

The original Hofstadter model considers spinless fermions confined to a two-dimensional square lattice potential and subject to a magnetic field in $z$ direction perpendicular to the plane. It describes electrons in a magnetic field giving rise to the (integer) Quantum Hall Effect (QHE). The single-particle spectrum as a function of the magnetic field strength was first computed by D. Hofstadter and exhibits the shape of a butterfly~\cite{hofstadter_energy_1976}. For a rational magnetic flux per plaquette $\alpha = p/q $, in units of the Dirac flux quantum $\Phi_0 = h/e$ and with coprime integers $p,q$, the system remains translationally invariant with an enlarged unit cell of $q$ lattice sites. All bulk bands carry non-zero Chern number $C_i$, and in all gaps one finds chiral edge modes. By the bulk-boundary correspondence, their total number is given by the sum of the Chern numbers of the filled bands below the gap: $N_c = \sum_{i \; \text{filled}} C_i$~\cite{thouless_quantized_1982, hasan_colloquium:_2010}. For even values of $q$ the middle bulk bands touch each other $q$ times at zero energy, where they form linearly dispersing Dirac cones~\cite{Wen1989641, PhysRevB.39.11943}.

In contrast, here we consider the time-reversal-invariant Hofstadter-Hubbard model of spinful and interacting fermions. To restore time-reversal symmetry we imagine applying a magnetic field in the $z$ direction that only couples to the $\ket{\uparrow}$-spins. A second field of the same strength that only couples to the $\ket{\downarrow}$-spins is applied in the opposite ($-z$) direction. We thus obtain a spinful and time-reversal-invariant version of the fundamental Hofstadter problem~\cite{PhysRevLett.105.255302}. While such a scenario is certainly not feasible with a ``real'' magnetic field, remarkably it can be experimentally realized using artificial gauge fields in cold atoms. The semi-metal (SM) at even values of $q$ now becomes a generalization of graphene with $q$ Dirac cones. The number of Dirac cones can be tuned by the artificial magnetic field strength and may become large. Moreover, the energy gaps which were crossed by $N_c$ chiral edge modes per edge in the QHE setup are now traversed by $N_h = N_c$ Kramer's pairs of helical edge modes per edge. The case of odd $N_h$ thus corresponds to the topological Quantum Spin Hall (QSH) phase, while an even number of $N_h$ corresponds to a normal insulator (NI) phase~\cite{PhysRevLett.95.146802, hasan_colloquium:_2010}. 

Particles with opposite spin may occupy the same lattice site and interact with each other. The interaction in cold-atom systems is short-ranged. We thus consider a Hubbard on-site interaction term in the Hamiltonian. The interaction strength may be tuned by Feshbach resonances or by adjusting the lattice depth. Apart from the time-reversal magnetic field and the Hubbard interaction term, there are additional terms available in the cold-atom experiment~\cite{PhysRevLett.105.255302,gerbier_gauge_2010}, such as a staggered optical lattice potential $\lambda_x$ and a Rashba-like spin-orbit coupling term $\gamma$ that breaks the axial spin symmetry as it induces spin flips between states $\ket{\uparrow}$ and $\ket{\downarrow}$ when the particle tunnels from site to site. 

The Hamiltonian of the time-reversal-invariant Hofstadter-Hubbard model with the additional terms takes the form
\begin{align}
  \label{eq:1}
  H = \sum_{j} \Bigl\{ -\bigl[ t_x c^\dag_{j + \hat{\bfx}} e^{- i 2 \pi \gamma \sigma^x} c_j + t_y c^\dag_{j + \hat{\bfy}} e^{i 2\pi \alpha x \sigma^z} c_j + \text{H.c.} \bigr] + (-1)^x \lambda_x c^\dag_j c_j +U n_{j, \uparrow} n_{j,\downarrow} \Bigr\} \,.
\end{align}
Here, $c_j^\dag=(c^\dag_{j\up}, c^\dag_{j\dw})$ is a spinor of the fermionic creation operators at lattice site $j = (x,y)$ with integer $x,y$, $\hat{\bfx}=(1,0)$, $\hat{\bfy}=(0,1)$ are unit vectors of the square lattice and $\sigma^{x,z}$ are the usual Pauli matrices. The parameters $t_{x}$ ($t_y$) denote the hopping amplitude in the $x$--($y$--)~direction. We focus on isotropic hopping $t_x = t_y = t$ here, and express all energies in units of $t \equiv 1$. The value of $\alpha $ determines the strength of the (artificial) magnetic field for either spin species which penetrates a lattice plaquette in units of the Dirac flux quantum. The on-site interaction strength is denoted by $U$, the parameter $\lambda_x > 0$ introduces a staggering of the optical lattice potential along the $x$--direction, and the continuous parameter $\gamma$ controls the relative amplitude of normal and spin-flip tunneling in the $x$--direction. 
Non-zero $\gamma$ breaks the axial spin symmetry and introduces mixing between the spin states $\{ \ket{\uparrow}, \ket{\downarrow}\}$. Since $\bigl[\sigma^x, \sigma^z \bigr] \neq 0$ the applied gauge field configuration is non-Abelian if both $\alpha, \gamma > 0$. For $\gamma = n/2 $ with $n \in \mathbb{Z}$ the part that is off-diagonal in spin space, which is proportional to $\sin 2 \pi \gamma$, vanishes. We may thus restrict $\gamma$ to the interval $[0, 1/2]$. 
For $U=0$ and $\alpha = 1/6$ the model in Eq.~(\ref{eq:1}) was studied in Ref.~\cite{PhysRevLett.105.255302}.


The remainder of the paper is structured as follows: in Sec.~\ref{sec:non-inter-phase}, we first study the energy spectrum and the phase diagram of the non-interacting system as a function of the parameters $\alpha, \gamma, \lambda_x$. We present the equivalent of the Hofstadter butterfly for non-zero spin-mixing $\gamma >0$ and staggering $\lambda_x > 0$. Then in Sec.~\ref{sec:interaction-effect}, we investigate the role of the Hubbard interaction in detail using RDMFT. We present the interacting phase diagram in Sec.~\ref{sec:inter-phase-diagr}, and explicitly determine the robustness of the topological phase with respect to the Hubbard-$U$. In Sec.~\ref{sec:inter-induc-topol}, we show that interactions may drive a transition from a trivial to a topological phase. Finally in Sec.~\ref{sec:tunable-magnetism}, we discuss the magnetic order exhibited by the system at large interactions and half filling. The magnetic phase diagram we observe is rich, containing various quantum phase transitions due to the TR-invariant magnetic field term $\alpha$ and the spin-mixing term $\gamma$.
Both terms may be tuned experimentally. Apart from N\'eel and collinear magnetic order we find various spiral phases. To understand the magnetic phase diagram we analytically derive the effective quantum Heisenberg Hamiltonian in this regime. We analyze this Hamiltonian in the classical limit using Monte-Carlo simulations and recover the same phases that we found in our RDMFT analysis of the itinerant model at large $U$. We end with a conclusion in Sec.~\ref{sec:conclusion}.

\section{Non-interacting phase diagram: spectrum and non-Abelian Hofstadter butterfly}
\label{sec:non-inter-phase}
\begin{figure}[t!]
  \centering
  \includegraphics[width=\linewidth]{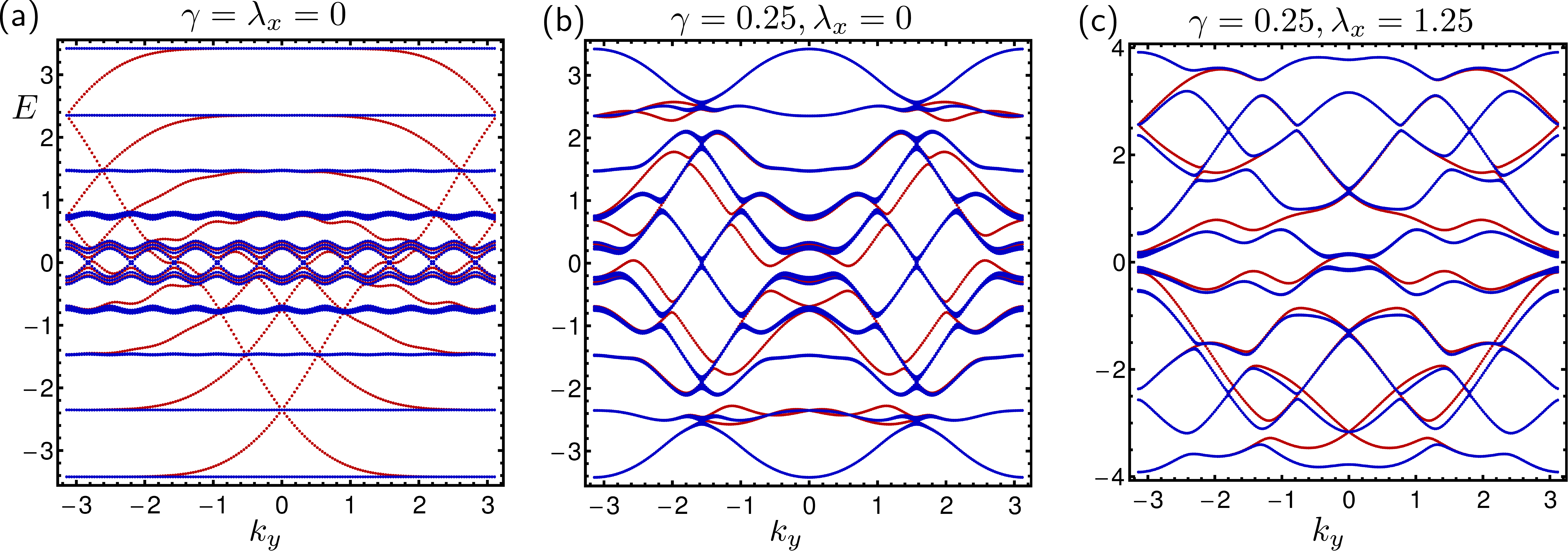}
  \caption{Energy spectrum for a cylinder geometry as a function of $k_y$. Bulk spectrum is shown in blue, edge states in red. Artificial magnetic flux is $\alpha = 1/10$, and $\gamma$ and $\lambda_x$ take values as indicated. Doubly degenerate helical edge states (red) are traversing the bulk gaps. }
  \label{fig:1}
\end{figure}

In this section we consider the non-interacting system and set $U = 0$. For $\gamma = 0$, the two spin species decouple and we obtain two time-reversed copies of the Hofstadter problem~\cite{hofstadter_energy_1976}. Fermions acquire a magnetic phase that depends on $x$ when tunneling along the $y$--direction. Opposite spins acquire opposite phases due to the factor of $\sigma^z$. The system remains periodic on the lattice if we choose the magnetic flux per plaquette to be $\alpha = p/q$ with coprime integers $p,q$. The unit cell contains $q$ sites, and consequently the magnetic Brillouin zone (MBZ) extends between $(k_x, k_y) \in [-\frac{\pi}{q}, \frac{\pi}{q}) \times [-\pi, \pi)$. The energy spectrum consists of $q$ bands. The bands are separated by finite energy gaps, except for the two middle bands in case of even $q$, which touch each other at $q$ distinct wave-vectors in the MBZ, where they form Dirac cones. Specifically, this occurs at $\bfk = \bigl(-\frac{\pi}{q}, \frac{\pm (2m + 1)\pi}{q} \bigr)$ for $q = 2 + 4 n$ and at $\bfk = \bigl( 0, \frac{\pm 2 m \pi}{q} \bigr)$ for $q = 4 + 4 n$ with $n,m \in \{0,1,\ldots\}$. Considering periodic boundary conditions (PBC) along the $y$--direction and open boundary conditions (OBC) along $x$, \emph{i.e.}, for cylindrical geometry, we present the energy spectrum for $\alpha = 1/10$ in Fig.~\ref{fig:1}(a). Within the gaps of the bulk spectrum (blue), we find a varying number of helical edge modes (red), specifically $N_h = \{1,2,3,4,4,3,2,1\}$ from bottom to top. This can be understood from an analogy to the corresponding QHE problem where the time-reversal invariant magnetic field is replaced by a real magnetic field, which breaks time-reversal symmetry. There, each bulk band carries a Chern number of $C=1$, except the two touching bands around zero energy which in total carry a Chern number of $C=-8$. Indeed, the number of chiral edge modes per edge is given by the sum over the Chern numbers of the filled bands $N_c = \sum_{i \; \text{filled}} C_i$. The sign of $N_c$ denotes the direction of the chiral edge mode. In presence of time-reversal symmetry, the sum over the Chern numbers for each gap of course vanishes, since up and down spins move in opposite directions and thus $\sum_{i \; \text{filled}} C^{\ket{\uparrow}}_i = - \sum_{i \; \text{filled}} C^{\ket{\downarrow}}_i$. 

The bulk spectrum in Fig.~\ref{fig:1}(a) is invariant under translation $k_y \rightarrow k_y' = k_y + 2 \pi/q$. This follows from the fact that for $\gamma = \lambda_x = 0$ the Hamiltonian $H$ in Eq.~\eqref{eq:1} commutes with the magnetic translation operator $T_x$, which is defined by $T_x c_j = \omega^{-y \sigma^z} c_{j + \hat{\bfx}} T_x $ with $\omega = \exp (i 2 \pi \alpha)$~\cite{zak_magnetic_1964,PhysRevA.83.013612}. The phase factor $\omega^{-y \sigma^z}$ accounts for the gauge field that breaks translational symmetry along $x$. The magnetic translation operator $T_y$ remains $T_y c_j = c_{j + \hat{\bfy}} T_y$. Note that while $[H,T_{x}] = [H,T_{y}] =0$, if $\gamma = \lambda_x = 0$, the translation operators do not commute between themselves $[T_x, T_y] \neq 0$, but only $[T_x^q, T_y ] = 0$. From $T_x c_{k_x, k_y} = c_{k_x, k_y + 2 \pi \alpha} T_x$ follows the additional degeneracy of the bulk spectrum (see Fig.~\ref{fig:1}(a)).

Let us now investigate the effect of non-zero $\gamma, \lambda_x \neq 0$ on the spectrum. If either of the two is non-zero, the Hamiltonian does not commute with $T_x$ anymore. While $H_{\lambda_x} = \sum_j (-1)^x \lambda_x c^\dag_j c_j$ still commutes with even powers of $T_x$, \emph{i.e.}, $\bigl[H_{\lambda_x}, T_x^{2n} \bigr] = 0$ for $n \in \mathbb{N}$, one finds that the $t_x$-hopping term only commutes with $T_x^{q/2}$ for $0 < \gamma < 0.5$. This follows directly from 
\begin{align}
  \label{eq:2}
  \sum_j \bigl[ T_x , c^\dag_{j+\hat{\bfx}} e^{- i 2 \pi \gamma \sigma^x} c_j \bigr] = \sum_j c^\dag_{j+\hat{\bfx}} i \sigma^x \sin (2 \pi \gamma) \Bigl[ \cos(4 \pi \alpha y) - 1 - i \sigma^z \sin (4 \pi \alpha y) \Bigr] c_j  T_x \,.
\end{align}
As an example, we present results for $\alpha = 1/10$ with non-zero $\gamma = 0.25$ and $\lambda_x = 0$ in Fig.~\ref{fig:1}(b). We observe that the bulk spectrum is still invariant under the shift $k_y \rightarrow k_y + 2 \pi \alpha \frac{q}{2} = k_y + \pi$. In contrast, if both $\gamma, \lambda_x \neq 0$ as in Fig.~\ref{fig:1}(c), the bulk spectrum has lost any translational symmetry. Note that it is still doubly degenerate at each $\bfk$-point and symmetric under $(k_x, k_y) \rightarrow (-k_x, - k_y)$ which follows from time-reversal invariance and inversion symmetry. 
\begin{figure}[t!]
  \centering
  \includegraphics[width=\linewidth]{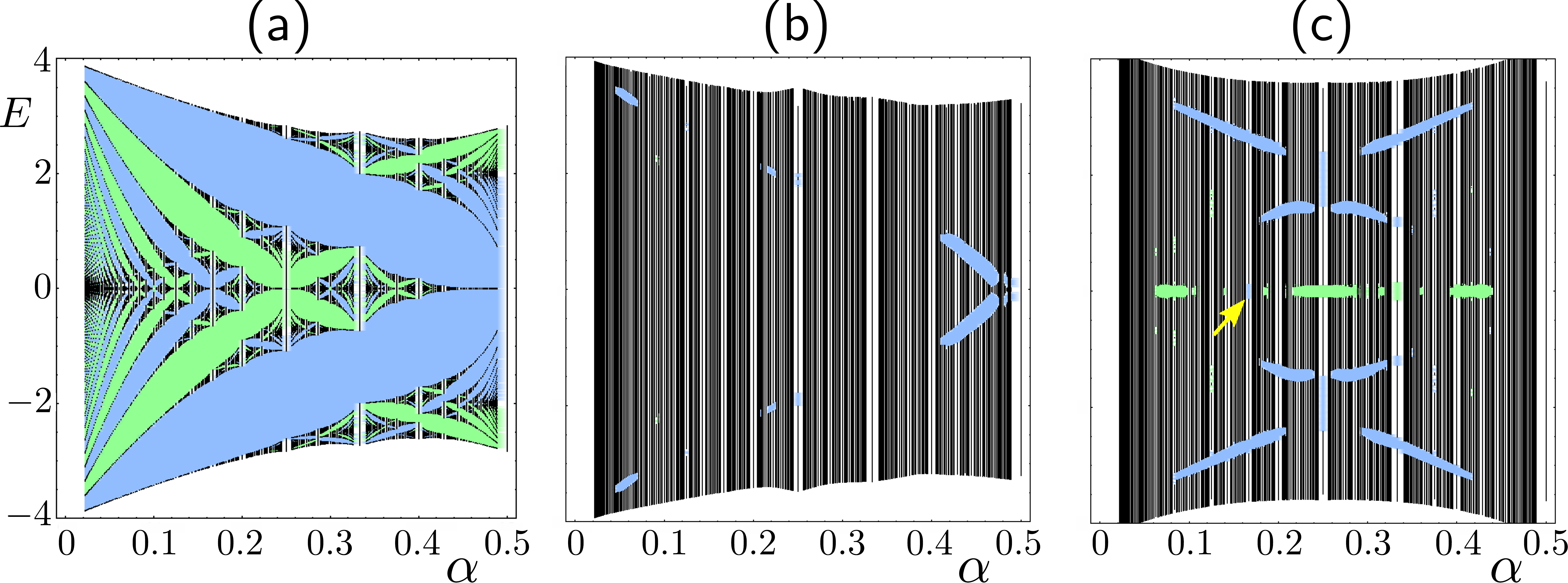}
  \caption{Hofstadter butterfly: energy spectrum (black) as a function of magnetic flux $\alpha = p/q$. We sample 500 different rational flux values $\alpha = \{1/2, 1/3, \ldots, 14/46, 15/46\}$. In (b) and (c) we observe that the system is predominantly metallic. The stripy appearance is due to the fact that we take $\alpha$ to be a rational number. The color inside the gap denotes the nature of the gapped phase: QSH (blue) and NI (green). Panel (a) is for $\gamma = \lambda_x = 0$, (b) for $\gamma = 0.1$, $\lambda_x =0.5$, and (c) is for $\gamma = 0.25$ and $\lambda_x = 1.0$. Note a QSH phase (blue) that appears at $E_F= 0$ for $\alpha = 1/6$ in panel (c) indicated by a yellow arrow. }
  \label{fig:2}
\end{figure}

The non-interacting system can thus be either (semi-)metallic, if the Fermi energy $E_F$ lies within a band or at a Dirac point, or it can be insulating, if $E_F$ lies within a gap. The normal and topological insulator are distinguished by a $\mathbb{Z}_2$ index $\nu$ which is equal to $\nu = N_h \; \text{mod} \;2$, where $N_h$ is the number of helical Kramer's pairs per edge. While we compute the $\mathbb{Z}_2$ index by counting the number of edge states for the interacting system, we use the efficient method introduced by Fukui and Hatsugai~\cite{PhysRevB.75.121403} in case of $U=0$. 

\begin{figure}[tbh]
  \centering
  \includegraphics[width=.7\linewidth]{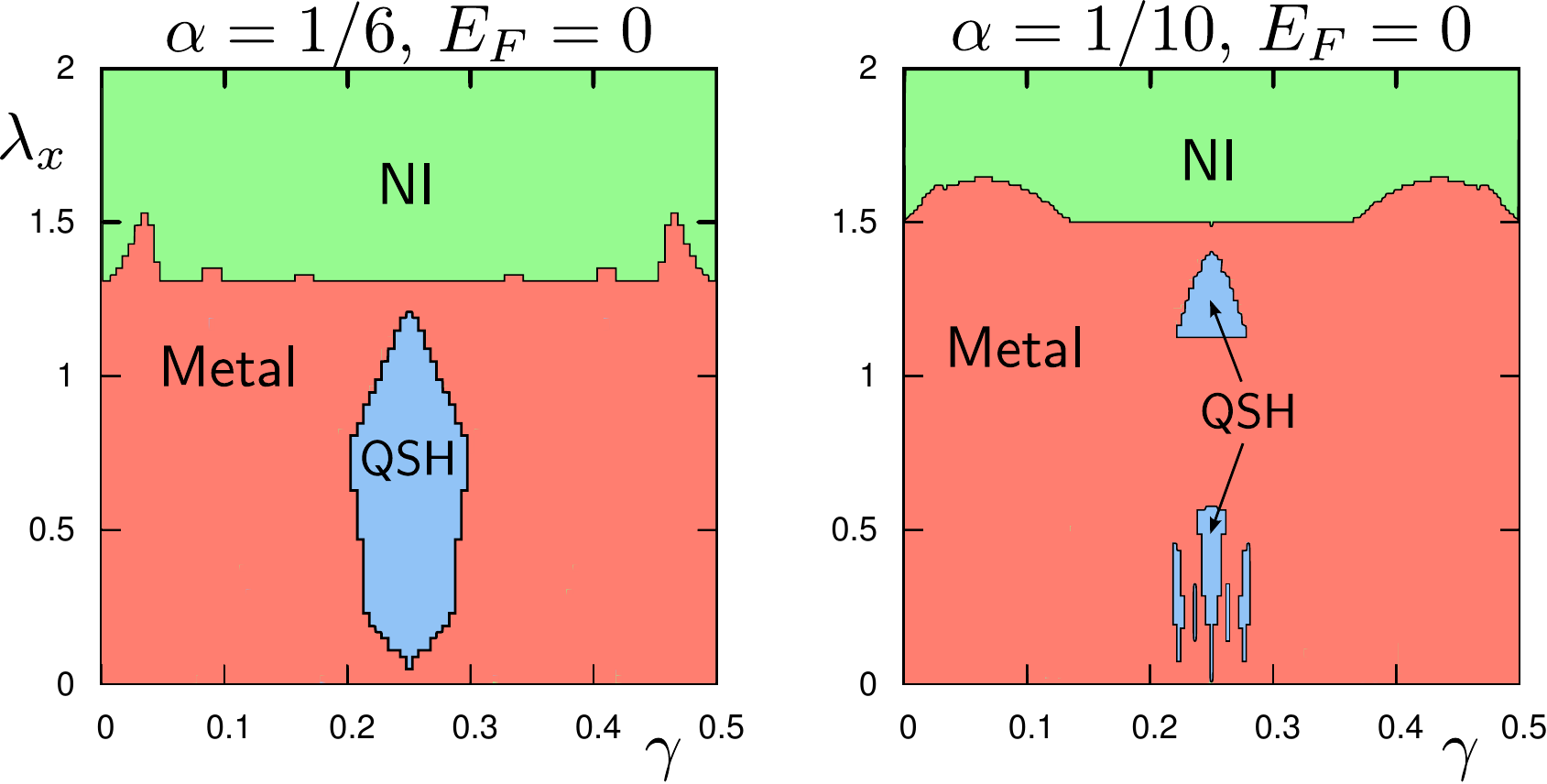}
  \caption{Non-interacting phase diagram for $\alpha = 1/6$ and $\alpha = 1/10$ at half-filling $E_F = 0$ as a function of $\gamma$ and $\lambda_x$. The QSH effect phase appears symmetrically around maximally spin-mixing $\gamma = 0.25$. For our finite size calculations, we identify the metallic regions when the ``gap'' size is smaller than the level spacing. Region of QSH effect and gap sizes are larger for $\alpha = 1/6$ than for $\alpha = 1/10$.}
  \label{fig:3}
\end{figure}

In Fig.~\ref{fig:2} we present the non-Abelian generalization of the Hofstadter butterfly for various values of $\gamma$ and $\lambda_x$. In Fig.~\ref{fig:2}(a), we show the familiar Hofstadter butterfly for $\gamma=\lambda_x = 0$, which shows the spectrum, \ie the bands, (in black) as a function of flux $\alpha = p/q$. Within the gaps, we have computed the $\mathbb{Z}_2$ index~\cite{PhysRevB.75.121403} and we denote topological (normal) insulator by blue (green). In Fig.~\ref{fig:2}(b), we present the non-Abelian butterfly for $\gamma = 0.1$ and $\lambda_x = 0.5$. Here, the translational degeneracy of the spectrum is lifted and the system is metallic in most parts of the energy range (see also Fig.~\ref{fig:1}(b-c)). Some gaps with $\nu=1$ have remained, however, for $\alpha \lesssim 1/2$. In Fig.~\ref{fig:2}(c), we show the butterfly for maximal spin-mixing $\gamma = 0.25$ and $\lambda_x = 1$. New gaps have opened, and in particular, we now find a TI phase at half-filling $E_F =0$ for $\alpha = 1/6$~\cite{PhysRevLett.105.255302}. Similarly, we find $\nu = 1$ for $\alpha = 1/10$ and slightly larger (and smaller) values of $\lambda_x$. To demonstrate, we choose two values of $\alpha =1/6$ and $\alpha=1/10$ and show the full $\lambda_x$--$\gamma$ phase diagram at half-filling in Fig.~\ref{fig:3}. The QSH phase appears around maximally spin mixing. Since the gap size is larger for $\alpha = 1/6$ this value is experimentally preferable. The case of half-filling is particularly interesting, since interactions drive a transition from a topological to an insulating state with exotic magnetic order, as we discuss in Sec.~\ref{sec:interaction-effect}.

Although we have found the QSH phase at half-filling only in the presence of spin-mixing $\gamma > 0$, it is not a requirement to break axial spin symmetry~\cite{schmidt-12prl156402,reuther-12prb155127}. Another way of opening the gap and inducing a QSH phase at half-filling \emph{without} breaking axial spin symmetry is to add a diagonal hopping term of the form $H_d = - t_d \sum_j \bigl( c^\dag_{j + \hat{\bfx} + \hat{\bfy}} e^{i 2 \pi \alpha \sigma^z (x + 1/2)} c_j + c^\dag_{j - \hat{\bfx} - \hat{\bfy}} e^{i 2 \pi \alpha \sigma^z (x - 1/2)} c_j + \text{H.c.} \bigr)$~\cite{PhysRevB.42.8282}. Our choice of Rashba-type hopping $\gamma$ and staggering $\lambda_x$ is thus only an experimentally motivated way~\cite{PhysRevLett.105.255302} of finding a QSH phase at half-filling. 


\section{Interaction effects: correlated topological phases and exotic magnetism}
\label{sec:interaction-effect}
One of the key advantages of simulating Hamiltonians with cold atoms in optical lattices is the ability to control the strength of the interaction via Feshbach resonances. 
Strong interactions and topologically non-trivial band structures can both give rise to novel phases of matter. In particular, when interactions and topological band structures are simultaneously present, unexpected exotic phases might arise, and novel phenomena emerge due to the interplay of topology and electron--electron interactions. Recently, the effect of on-site interactions on topological band structures has been extensively studied~\cite{rachel-10prb075106,PhysRevLett.106.100403, PhysRevB.84.205121, PhysRevLett.107.010401, PhysRevLett.107.166806,PhysRevB.85.205102,PhysRevLett.108.046401,PhysRevB.85.045123}; for details we refer the reader to a recent review article~\cite{hohenadler-12arXiv:1211.1774}. 

Here, we shed light on this issue using RDMFT and thoroughly study the effect of interactions on the band structure discussed above. We explicitly show the robustness of the edge states up to relatively strong interactions. Further, we identify a scenario where the interactions drive the system from a normal to a topological phase. At half-filling, the system exhibits magnetic order at large $U$, and we find the magnetic phase diagram using RDMFT. We also rigorously derive a quantum Heisenberg Hamiltonian in this regime, and investigate it numerically in the classical limit of large spin $S$.


\subsection{Interacting phase diagram and stability of topological phases}
\label{sec:inter-phase-diagr}
Let us now study the influence of on-site interaction on the phase diagram. We use RDMFT to determine the spectral function $A(x,k_y, \omega)$ in a cylinder geometry with PBC along $y$ and OBC along $x$. This allows us to directly map out and count the helical edge states, which determines the $\mathbb{Z}_2$ index $\nu$. Edge states are also used experimentally to measure $\nu$~\cite{PhysRevLett.105.255302,GoldmanGerbier_DetectingEdgeStates-arXiv_2012,PhysRevA.85.063614, goldman-arXiv:1212.5093-2012}. In the following we focus on fixed $\alpha = 1/6$, where all relevant phenomena that occur in this system for general $p/q$ are qualitatively captured. 

Away from half-filling, we have investigated the stability of the topological phases. Placing the Fermi energy in the lowest gap at filling factor $n_F = 1/3$ fermions per site, we find that within RDMFT these phases appear to be robust with respect to interactions. We have computed the spectral function for a finite size system using RDMFT and the helical edge mode is clearly visible even for $U=10$. 

\begin{figure}[tbh]
  \centering
  \includegraphics[width=\linewidth]{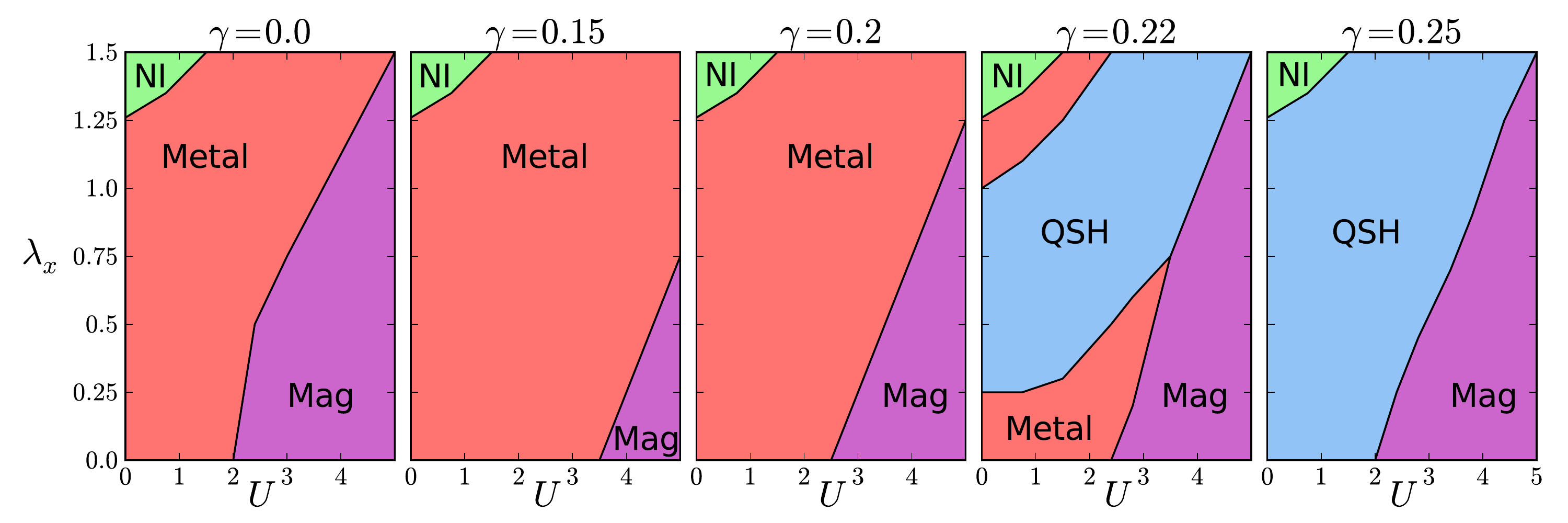}
  \caption{Interacting phase diagram at $\alpha = 1/6$ as a function of interaction $U$ and staggering $\lambda_x$ for various values of spin-mixing $\gamma$. We find (semi)-metallic regime (Metal), magnetically ordered state (Mag), normal insulator (NI) and quantum spin Hall state (QSH).  }
  \label{fig:5}
\end{figure}

At half-filling, we have determined the interacting phase diagram for $\alpha = 1/6$ as a function of $\lambda_x$ and $U$ for various values of $\gamma$, and it is shown in Fig.~\ref{fig:5}. Since the phase diagram is symmetric around $\gamma = 0.25$ is it sufficient to consider $0 \leq \gamma \leq 0.25$. At $U=0$ the system is a (semi)-metal for $\gamma < 0.2$ and $\lambda_x \lesssim 1.26$. It turns into a normal band insulator for $\lambda_x \gtrsim 1.26$ (see Fig.~\ref{fig:3}~(left)). At large $U \geq U_c(\gamma, \lambda_x)$ the system exhibits magnetic order. The type of magnetic order depends on $\gamma$, as we show in detail in Sec.~\ref{sec:tunable-magnetism}. The critical $U$ increases with $\lambda_x$, which follows from the fact that $\lambda_x$ tends to localize fermions in every other column whereas $U$ favors a state with exactly one fermion per site. In general, increasing the interaction strength $U$ reverses the effect of the staggering potential $\lambda_x$. This leads to the interesting situation at $0.2 < \gamma < 0.3$, where we find interaction driven transitions from a metallic or a normal insulating phase into the topological QSH phase. The QSH phase then extends over a broad range of about $U \approx 2-4$ before the system eventually develops magnetic order at $U = 4-5$ and becomes gapped again. 

\subsection{Interaction induced topological quantum phase transition}
\label{sec:inter-induc-topol}
The non-trivial topology of topological insulators is an effect entirely due to the band structure, \ie the topology is already contained in the non-interacting part of the Hamiltonian. On the other hand, it was realized that topological band properties can also arise in topologically trivial systems from the spontaneous breaking of a symmetry due to interactions~\cite{raghu-08prl156401}. This idea suggests that topological insulators could be present in a much larger class of materials. Recently a few theoretical models have been proposed where this scenario is realized~\cite{PhysRevLett.109.205303,raghu-08prl156401,wang-12epl57001,budich-12prb201407,sun-09prl046811,uebelacker-11prb205122,liu-10prb045102,zhang-09prb245331,dzero-12prb045130, PhysRevB.84.201103, PhysRevB.85.235135}; from the experimental point-of-view, the situation is, however, less satisfying and an experimental study of an interaction-driven topological insulator is still lacking. 

In principle, one should distinguish at least two different scenarios: in the first case the non-interacting band structure does not exhibit any topological phase like QSH or Chern insulator, and interactions are solely responsible for inducing the topological phase (see \emph{e.g.} Ref.~\cite{raghu-08prl156401}). In the second case the non-interacting band structure does in principle contain a QSH or Chern insulator phase but the parameters are tuned such that the model is in its trivial phase. Then the interactions cause a renormalization of the band structure or effective masses driving the system into the topological phase. 

A simple example of an interaction-induced topological quantum phase transition into a QSH phase, which belongs to the second class, is realized in the Hamiltonian~\eqref{eq:1} for $\alpha=1/6$. Here we thus provide an example which is experimentally relevant. We will discuss this topological quantum phase transition in the following.

In principle, interactions are encoded in the self energy $\Sigma(\boldsymbol{k},\omega)$. In most approaches, one needs to approximate the self energy for practical reasons. Usually one uses approximations where the self energy becomes either purely momentum-dependent or purely frequency-dependent. 
\begin{figure}[t!]
  \centering
  \includegraphics[width=\linewidth]{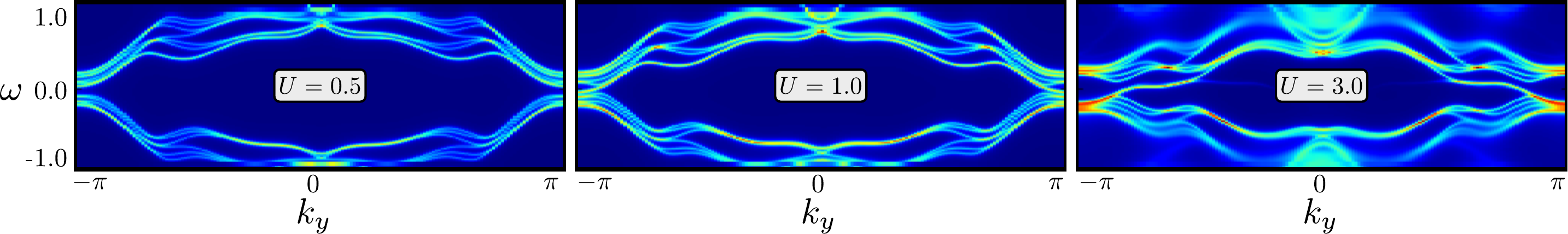}
  \caption{Intensity plot of the single-particle spectral function $A(k_y,\omega) = \int_{k_x} A(k_x, k_y, \omega)$ as a function of $\omega$ and $k_y$ for $\alpha=1/6$, $\lambda_x=1.5$, $\gamma=0.25$ and different interaction strength $U$. Left panel is for $U=0.5$, where system is in the normal insulating phase, (middle) is for $U = 1.0$ where we find a metallic phase, and (right) is for $U = 3.0$, where the system is in the quantum spin Hall phase. The spectral function visualizes the topological difference, in (left) no edge state is crossing the bulk gap while in (right) a pair of helical edge states is traversing the bulk gap.}
  \label{fig:akw}
\end{figure}
In static mean-field, for instance, the self energy is reduced to $\Sigma(\boldsymbol{k})$  (or even to a constant $\Sigma_0$) and enters as a band structure renormalization. In certain cases, such a band structure renormalization can cause band-inversion and induces the topological phase. Another proposal discussed a fluctuation-induced topological phase transition~\cite{budich-12prb201407, PhysRevB.85.235135}, \ie a topological phase induced solely by $\Sigma(\omega)$ being momentum-independent. 

In our Hamiltonian~\eqref{eq:1}, the topological phase transition can be understood as a $\Sigma_0$-driven transition. This constant part of the self energy is contained as a lowest order contribution in both $\Sigma(\bfk)$ and $\Sigma(\omega)$. The latter quantity is available in DMFT-like methods and thus we can observe this interaction-induced topological phase transition. 
The transition appears for parameters such as $\gamma=0.22$ and $\lambda_x=1.5$ where the band structure is in a topologically trivial phase. The reader should note that large values of the staggered potential $\lambda_x$ avoid the QSH phase, \ie for the same parameters but $\lambda_x = 0.75$ the QSH phase would be present. By increasing $U$ we effectively reduce the effect of $\lambda_x$ and drive the system into a metallic phase. Under further increase of $U$ the gap re-opens and one enters the QSH phase. Consequently, these transitions from trivial insulator to metal and from metal to QSH phases are due to interactions which cause a band structure renormalization. We have shown the corresponding spectral functions in Fig.~\ref{fig:akw} for the three different regimes. Absence or presence of edge states, respectively, distinguish the trivial and topological insulator phases.


\subsection{Tunable exotic magnetism}
\label{sec:tunable-magnetism}
We learned previously that at half-filling, the system turns into a magnetically ordered insulator at $\alpha = 1/6$ and large interaction strength $U \gg t$ (see Fig.~\ref{fig:5}). This behavior occurs generally in the half-filled TR-invariant-Hofstadter-Hubbard model at large $U$. Exotic magnetic order appears that depends on the value of the Rashba-type spin-orbit coupling $\gamma$ and the artificial magnetic flux $\alpha$. Importantly, by tuning these parameters one crosses various magnetic quantum phase transitions from N\'eel order to various spiral ordered phases to collinear order. Similar phases have been reported at strong interactions in other cold-atom setups that include non-Abelian gauge fields~\cite{PhysRevLett.109.085302, PhysRevLett.109.085303, PhysRevA.85.061605, gong-arxiv-2012}. 

\begin{figure}[t!]
  \centering
  \includegraphics[width=\linewidth]{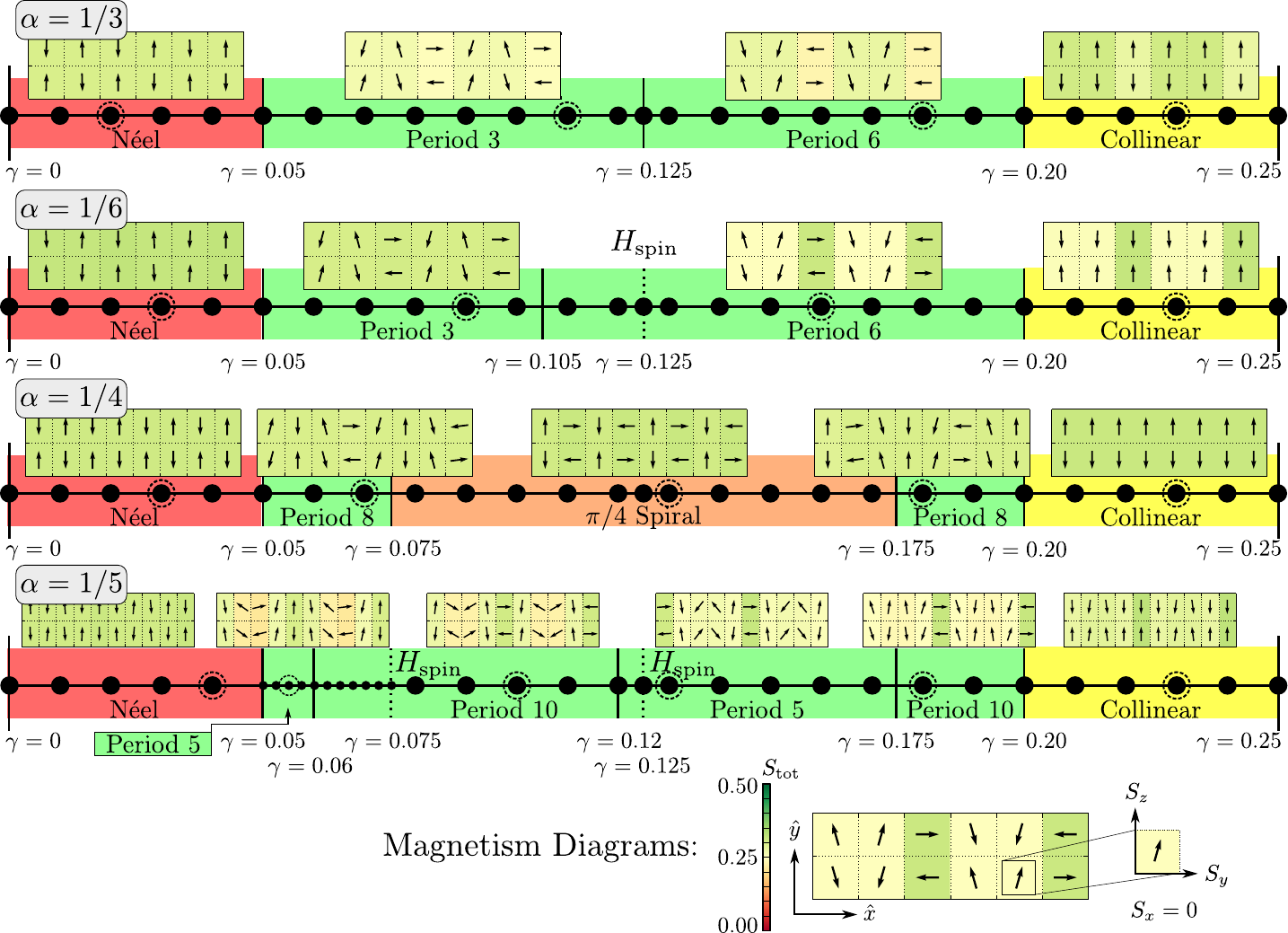}
  \caption{Magnetic order of the itinerant model at $U = 5$ as a function of $\gamma \in [0, 0.25] $ for different values of $\alpha = 1/3, 1/6, 1/4, 1/5$ (from top to bottom) obtained within RDMFT. We generally find N\'eel order around $\gamma =0$ and collinear order around $\gamma = 0.25$. In between for $0.05 \lesssim \gamma \lesssim 0.20$, we observe different kinds of spiral order with periodicities of three, four, six, and eight sites due to a competition of a Dzyaloshinskii-Moriya (DM) and an anisotropic XXZ N\'eel type spin interaction, see Eq.~\eqref{eq:3}. Phase boundaries are indicated by solid vertical lines. Black dots denote the actual values of $\gamma$ that we simulated using RDMFT. The magnetization diagrams show the direction of magnetic order in $S_y$--$S_z$--plane. The background color indicates the magnetization amplitude $S_{\text{tot}}(x)$. The magnetization diagrams are extracted from the specific parameter point that is indicated by the dashed circle. If, due to quantum fluctuations, the classical spin Hamiltonian exhibits a different phase boundary than the itinerant model, we denote the phase boundary of the spin Hamiltonian by a vertical dashed line with label $H_{\text{spin}}$. }
  \label{fig:6}
\end{figure}

Using RDMFT we determine the magnetic phase diagram in Fig.~\ref{fig:6} as a function of $\gamma$ for various values of $\alpha$. We work at fixed interaction strength $U=5$. We complement our analysis by rigorously deriving a quantum spin Hamiltonian at small $t_{x,y}^2/U$ when charge fluctuations freeze out and we numerically determine its magnetic phase diagram in the classical limit of large $S$ using classical Monte-Carlo simulations, and compare to the RDMFT results of the itinerant model. While the phase boundaries are often slightly shifted, we generally find the same type of magnetic order in both the itinerant model at $U=5$ and the classical spin Hamiltonian. 

In Fig.~\ref{fig:6}, we observe that the order along the $y$ direction is always antiferromagnetic, and we find various types of order along the $x$ direction. In all cases investigated, the $S_x$ component is zero, and we can represent all spins in the $S_z-S_y$ plane. For $0 \leq \gamma \leq 0.05$ one finds usual N\'eel order for all four values of $\alpha$ that we consider. Around maximal spin-mixing $0.2 \leq \gamma \leq 0.25$ the order is collinear with spins ordering ferromagnetically along the $x$--direction and antiferromagnetically along the $y$--direction. In between, the system exhibits various forms of spiral order with different periods of three, four, five, six, eight or ten lattice sites for the values of $\alpha$ that we present here. 

We can describe the spiral states using a variational ansatz with only four parameters. The order of the spin Hamiltonian for each set of parameters is uniquely defined by the angle of rotation $\theta_i$ between nearest-neighbor spins along the $x$--direction, \emph{i.e.}, $\cos(\theta_i) = \bfss_i\cdot \bfss_{i+\hat{\bm{x}}}/|\bfss_i| |\bfss_{i+\hat{\bm{x}}}|$. A single ansatz of the form
\begin{equation}
\theta_i = \phi_0 + \phi_\omega \cos(\omega x + \eta)
\end{equation}
with the four variational parameters $\{\phi_0, \phi_\omega, \omega, \eta\}$ perfectly fits the order found. Note that $\theta_i = \phi_0$ is sufficient to describe N\'eel, collinear, and simple spiral order, which are indicated in Fig.~\ref{fig:6} by red, yellow and orange regions respectively. We distinguish these from the more complex spiral order, indicated by green regions, that can only be described with one additional Fourier component of frequency $\omega$. We determine these variational parameters by a Fourier analysis of the Monte-Carlo data and, as one might expect, these parameters are close to nice analytical values. For example, $\alpha=1/6$, $\gamma=0.15$ has the values $\phi_0\approx\pi/3$, $\omega\approx2\pi/3$, $\phi_\omega\approx\pi/6$.

\begin{figure}[tbh]
  \centering
  \includegraphics[width=.8\linewidth]{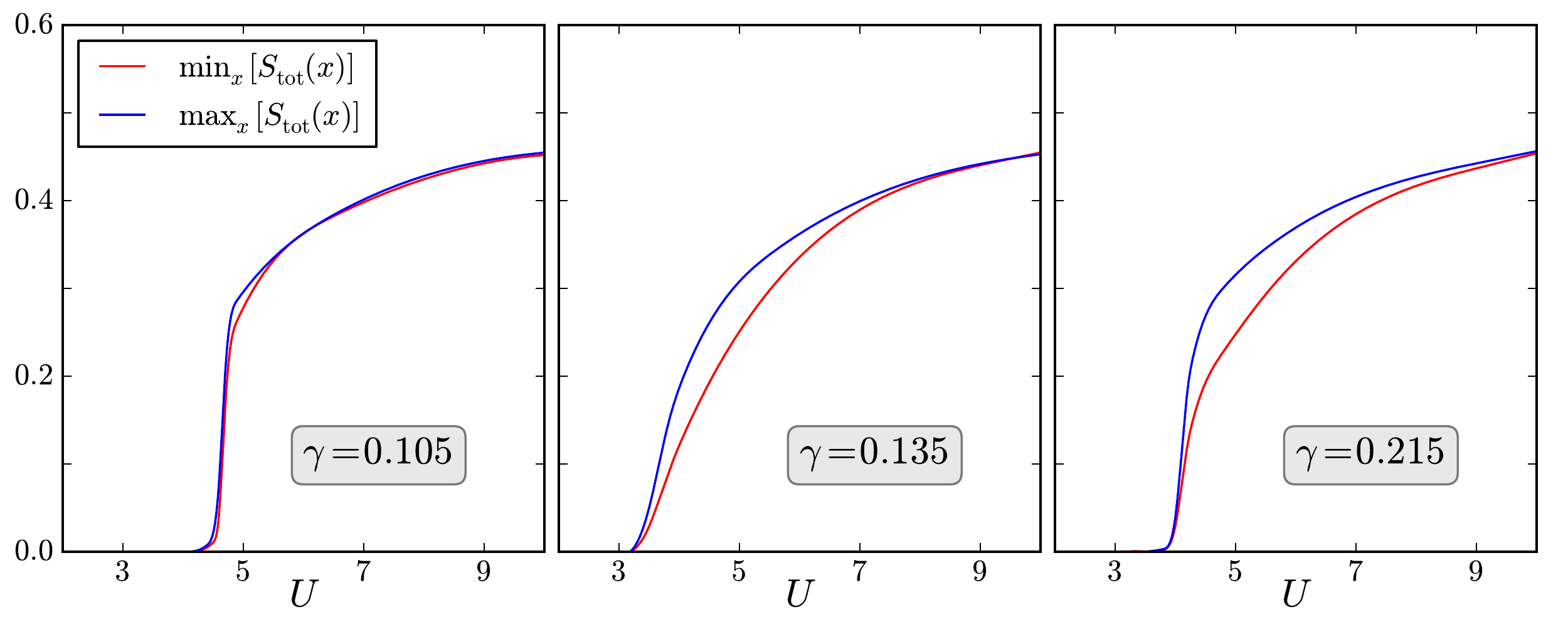}
  \caption{Spatial staggering of the magnetization amplitude $S_{\text{tot}}(x)$ as a function of $U$ for $\alpha = 1/6$. Staggering is defined by the difference of the maximal and the minimal $S_{\text{tot}}$ within one period of the spin order. The staggering is indicated by the colored background in the spin boxes of Fig.~\ref{fig:6}. We observe that the staggering decreases for increasing $U$, because the fermions become more and more localized and quantum fluctuations are suppressed. This also occurs for the other values of $\alpha$ that we investigated. }
  \label{fig:20}
\end{figure}

The magnetic phase diagram is symmetric around the point $\gamma = 0.25$, except that the spiral phases have different chirality for $0.3 \leq \gamma \leq 0.45$. 
Quantum fluctuations tend to reduce the amplitude of the magnetization $S_{\text{tot}} < 1/2$, which depends on the values of $\gamma$, $\alpha$ and $U$. In addition, for collinear and some spiral orders $S_{\text{tot}}(x)$ is also spatially staggered along the $x$ direction for intermediate interaction strength $U$, which we indicate by the colored background of the spin boxes in Fig.~\ref{fig:6}. As shown in Fig.~\ref{fig:20}, this spatial staggering reduces with increasing $U$ as the fermions become more and more localized.
The type of magnetic order that occurs can be understood by rigorously deriving a quantum spin Hamiltonian at large interactions when the charge degrees of freedom freeze out and only virtual hopping events take place~\cite{PhysRevLett.109.205303,auerbach_quantum_magnetism}
\begin{align}
  \label{eq:3}
& \mathcal{H} = J_x \sum_j \biggl\{  S^x_j S^x_{j+\hat{\bfx}} + \cos (4 \pi \gamma) \Bigl[S^y_{j} S^y_{j+\hat{\bfx}} + S^z_{j} S^z_{j+\hat{\bfx}} \Bigr] + \sin( 4 \pi \gamma ) \Bigl[ S^z_j S^y_{j+\hat{\bfx}} - S^y_{j} S^z_{j+\hat{\bfx}}  \Bigr] \biggr\} \nonumber \\ & + J_y \sum_{j} \biggl\{ \cos \bigl( 4 \pi \alpha x \bigr) \Bigl[ S^x_{j } S^x_{j+\hat{\bfy}} + S^y_{j} S^y_{j+\hat{\bfy}} \Bigr] + S^z_{j} S^z_{j+\hat{\bf y}} + \sin( 4 \pi \alpha x) \Bigl[ S^y_j S^x_{j + \hat{\bfy}} - S^x_j S^y_{j + \hat{\bfy}}  \Bigr] \biggr\} \,,
\end{align}
where $J_i=4t_i^2/U$ and the spin operators are defined in terms of the fermionic operators as $\bfss_j = \frac12 c^\dag_j \boldsymbol{\sigma} c_j$ with $\boldsymbol{\sigma}= (\sigma^x, \sigma^y, \sigma^z)$. The first part describes spin exchange in $x$ direction, which contains the Rashba-coupling $\gamma$.  For $\gamma = n/2$ with $n \in \mathbb{Z}$, the coupling takes the form of the familiar Heisenberg Hamiltonian. For other values of $\gamma$, however, the $SU(2)$ symmetry is broken down to $XXZ$-type with anisotropy direction $S^x$ in spin space. For $\gamma \neq n/4$ an additional Dzyaloshinskii-Moriya (DM) interaction term in the $YZ$ plane is present. It is this term that is responsible for the spiral order that we find for intermediate values of $\gamma$. In the absence of an artificial magnetic field ($\alpha=0$) no direction is preferred for spiral order (in $YZ$) or antiferromagnetic order (in $X$).

The second part of Eq.~\eqref{eq:3} describes spin exchange in $y$ direction, and depends on the artificial magnetic field strength $\alpha$. It is periodic along the $x$ direction with an extended unit cell of $q/2$ lattice sites. For $\alpha = 1/2$ one obtains the ordinary Heisenberg exchange, while for other values of $\alpha$ the spin interaction of the $S^x$ and $S^y$ components exhibit an $x$ dependent modulation of the variable strength of Heisenberg exchange and DM-type interaction. The spin interaction along $Z$ is always antiferromagnetic. In the absence of a Rashba-like term ($\gamma=0$) the periodic modulation causes ordering in the $X$ and $Y$ directions to be unfavorable, such that only $Z$ antiferromagnetic order occurs. For $\gamma \neq 0$, the competition between the present terms causes the $X$ direction to be unfavorable and only order in the $Y$--$Z$ plane is observed.

To make a connection between the spin model~\eqref{eq:3} and the itinerant model~\eqref{eq:1} at large $U$, we have solved for the phase diagram of the spin model in the classical limit of large spin $S$ using Monte-Carlo simulations. As shown in Fig.~\ref{fig:6}, we find that while the phase boundaries between the different magnetically ordered regions are shifted in some cases, which is not surprising considering that we are taking the classical limit of a $S=1/2$ quantum spin model, the type of magnetic order that occurs in the two models is the same. 

\section{Conclusions}
\label{sec:conclusion}
In conclusion, ultracold atoms in optical lattices in the presence of non-Abelian gauge fields provides a rich platform to study the influence of interactions on topological phases. Motivated by experiment, we have investigated a specific setup, the time-reversal-invariant Hofstadter-Hubbard model, and determined the interacting phase diagram. Interactions drive various phase transitions. We identified a scenario where a normal insulator becomes topologically non-trivial for increasing interactions. At large interactions and half-filling, the systems develops tunable exotic magnetic order with N\'eel, collinear, and spiral phases due to the interplay of Rashba-type spin-orbit coupling and the artificial magnetic field. We explicitly prove the robustness of helical edge states in the correlated topological phases, which is crucial for cold-atom detection schemes.

\ack
The authors acknowledge fruitful discussions with I. B. Spielman. The Young Investigator Group of P.P.O. received financial support from the ``Concept for the Future'' of the KIT within the framework of the German Excellence Initiative. S.R. acknowledges support by the DFG through FOR 960. K.L.H. acknowledges support from NSF via DMR-0803200. W.H. acknowledges financial support from the Deutsche Forschungsgemeinschaft (DFG) via Sonderforschungsbereich SFB/TR 49 and Forschergruppe FOR 801. 

\vspace{.8 cm}


\end{document}